\documentclass[twocolumn,preprintnumbers,superscriptaddress,nofootinbib,aps,prd,floatfix]{revtex4-2}

\usepackage{enumerate}
\usepackage{subfig}
\usepackage{amsmath}
\usepackage{amssymb}
\usepackage{graphicx}
\usepackage{xspace,slashed}
\usepackage{xcolor}
\usepackage{hyperref}
\hypersetup{colorlinks=true, citecolor=blue, urlcolor=blue, linkcolor=blue}
\usepackage{bbm}
\usepackage{float}
\usepackage{multirow}
\usepackage{orcidlink}

\newcommand{\sw}{s_W}
\newcommand{\cw}{c_W}

\newcommand{\U}{\boldsymbol{U}}

\begin{document}

\title{Higgs boson off-shell measurements probe non-linearities}
\author{Anisha\orcidlink{0000-0002-5294-3786}} \email{anisha@glasgow.ac.uk}
\affiliation{School of Physics \& Astronomy, University of Glasgow, Glasgow G12 8QQ, United Kingdom}
\author{Christoph Englert\orcidlink{0000-0003-2201-0667}}\email{christoph.englert@glasgow.ac.uk}
\affiliation{School of Physics \& Astronomy, University of Glasgow, Glasgow G12 8QQ, United Kingdom}
\author{Roman Kogler\orcidlink{0000-0002-5336-4399}}  \email{roman.kogler@desy.de}
\affiliation{Deutsches Elektronen-Synchrotron DESY, Notkestr. 85, 22607 Hamburg, Germany}
\author{Michael Spannowsky\orcidlink{0000-0002-8362-0576}} \email{michael.spannowsky@durham.ac.uk}
\affiliation{Institute for Particle Physics Phenomenology, Department of Physics, Durham University, Durham DH1 3LE, United Kingdom}

\begin{abstract}
The measurements of off-shell Higgs boson contributions in massive gauge boson pair production are known to probe its electroweak interactions across different energy scales. Often employed as an estimator of the Higgs boson width in restricted theories of beyond the Standard Model physics, we revisit this measurement and re-advertise its potential to constrain aspects of Higgs boson non-linearity. We show that this so-called off-shell measurement complements related analyses of multi-Higgs final~states.
\end{abstract}
\preprint{DESY-24-018}
\preprint{IPPP/24/01}
\maketitle
\allowdisplaybreaks
\section{Introduction}
\label{sec:intro}
Even some ten years after its discovery, the Higgs boson remains at the core of the experimental quest for new physics beyond the Standard Model (BSM). Given that searches for new states around the TeV scale have so far been unsuccessful, the methodology of Effective Field Theory (EFT) becomes increasingly relevant for the interpretation of Large Hadron Collider (LHC) data, alongside a theoretically useful framing of measurement uncertainties.  Most efforts along these lines have concentrated on the so-called Standard Model EFT (SMEFT), largely at the dimension-six level that constructs effective interactions from SM fields such as the Higgs doublet. As a consequence, SMEFT predicts strict correlations across Higgs multiplicities~\cite{Gomez-Ambrosio:2022why,Bhardwaj:2023ufl,Delgado:2023ynh}. This, by construction, reduces the qualitative relevance of multi-Higgs production modes as part of a global fit.

From this perspective, in the electroweak chiral Lagrangian (or non-linear Higgs EFT~\cite{Appelquist:1980vg,Longhitano:1980iz,Longhitano:1980tm,Feruglio:1992wf,Brivio:2013pma,Buchalla:2013rka,Buchalla:2013eza}, HEFT), any Higgs coupling can be considered a free parameter. On the one hand side, this leads to a significant growth of free parameters, which reduces the value of LHC data as cancellations between couplings naturally imply a loss of sensitivity. On the other hand, existing approaches to analyses pursued by the experimental community can only be interpreted in this framework when SM gauge-related couplings are treated as independent parameters, e.g., in the so-called $\kappa$ framework~\cite{LHCHiggsCrossSectionWorkingGroup:2011wcg}. Furthermore, current data still allows for considerable admixture of electroweak singlet states, and associated non-linearity should be measured and constrained, and not imposed in investigations parallel to the SMEFT programme.

How can we constrain such interactions efficiently in the future? Although the aforementioned multi-Higgs programme certainly is an avenue, given the comparably small production cross sections at the LHC, it might not provide a conclusive picture. To this end, we revisit the off-shell Higgs measurement $pp\to 4\ell$~\cite{Caola:2013yja} in the context of HEFT. We show that this process, which is usually framed from the perspective of top-Yukawa measurements correlated with the Higgs width under SM assumptions, provides significant power to constrain Higgs boson non-linearity. This is routed in tell-tale cancellations of related effects in the context of SMEFT, paired with the non-decoupling of the propagating Higgs contribution as a consequence of unitarity~\cite{Kauer:2012hd}.

This note is organised as follows: We highlight a particularly relevant set of interactions that enable the discrimination of SMEFT vs. HEFT from non-trivial momentum dependencies that are accessible as part of the off-shell Higgs contribution in, e.g., $pp\to H \to ZZ$ in Sec.~\ref{sec:linearEFT}. In Sec.~\ref{sec:chiralEFT}, we detail all relevant HEFT interactions and their relation to SMEFT, we also comment on details of our implementation. In Sec.~\ref{sec:offshell}, we discuss the constraints on Higgs non-linearity that the off-shell measurement can offer. We conclude in Sec.~\ref{sec:conc}.

 \section{Linear vs. non-linear momentum dependencies}
 \label{sec:linearEFT}
It is instructive to highlight a particular class of operators that transparently display the differences between HEFT and SMEFT we seek to capitalise on. In the SILH-like basis~\cite{Giudice:2007fh,Wells:2015uba,Englert:2019zmt,Henning:2014wua,Englert:2017aqb,Buchalla:2014eca} involving only bosonic operators, there is a dimension-6 CP-even operator of class $D^4 \Phi^2$ that gives rise to a quartic momentum dependence of the Higgs propagator
\begin{equation}
\label{eq:smeft}
	Q_{\square \Phi}= \frac{C_{\square \Phi}}{\Lambda^2}|D^{\mu} D_{\mu} \Phi |^2\,.
\end{equation}
Here, $\Phi$ is the SM $SU(2)_{L}$ scalar doublet and $D_\mu$ is the covariant derivative. In the broken phase, 
\begin{equation}
\Phi={1\over \sqrt 2} \left( \begin{matrix} 0 \\ {v+H }\end{matrix} \right)
\end{equation}
with $v\simeq 246~\text{GeV}$. Extending the SM Lagrangian with this dimension-6 operator, in the broken phase, modifies the Higgs two-point function. The corresponding vertex function is written as 
\begin{multline}\label{eq:HH_d6}
	\parbox{2.6cm}{\vspace{0.2cm}\includegraphics[width=2.6cm]{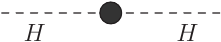}} \\=-i \Sigma(p^2)
	=i(p^2-m_{H}^2) +i \frac{C_{\square \Phi }}{\Lambda^2} p^4\,.
\end{multline}
To obtain the on-shell renormalised Higgs two-point function, the Higgs field is modified as $\sqrt{Z}H=(1+\delta Z_{H}/2)H$, where $Z_{H}$ is the Higgs wave function renormalisation in the on-shell scheme is given as
\begin{equation}
	\delta Z_{H}= \frac{ d\Sigma(p^2)}{dp^2}\Big|_{p^2=m_{H}^2 }= -\frac{2 C_{\square \Phi}}{\Lambda^2} m_{H}^2\,.
\end{equation}
Including these corrections, the Higgs propagator becomes~\cite{Englert:2019zmt}  
\begin{multline}
\label{eq:higgsprop}
	\Delta_{H}(p^2)= \frac{1}{p^2 -m_{H}^2}-\frac{C_{\square \Phi }}{\Lambda^2} \\
	= \frac{1}{p^2 -m_{H}^2}\Big(1-\frac{C_{\square \Phi }}{\Lambda^2}(p^2 -m_{H}^2)\Big)\,.
\end{multline}
Higher point functions also receive dimension-6 corrections and the corresponding Feynman rule for, e.g., the $HZZ$ three-point vertex function is
\begin{multline}\label{eq:HZZ_d6}
	\parbox{2.6cm}{\vspace{0.cm}\includegraphics[width=2.6cm]{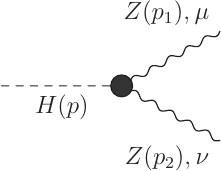}} = i \Gamma^{\mu\nu}_{HZZ}(p,p_1,p_2)\\
	=\frac{i e^2 v}{2 c_{_{W}}^2 s_{_{W}}^2}\Big[g^{\mu\nu} +\Big(
	p^2g^{\mu\nu}-p^{\nu}
	p_2^{\mu}-p_1^{\mu}p_2^{\nu} -p^{\mu} p_2^{\nu}\Big)\frac{C_{\square \Phi }}{\Lambda^2}\Big]\,.
\end{multline}
To gain a qualitative understanding of the overall amplitude modification induced by $Q_{\square \Phi}$ (neglecting all $1/\Lambda^4$ terms), we combine the Eqs.~\eqref{eq:HH_d6} and~\eqref{eq:HZZ_d6}
\begin{equation}
\label{eq:HZZ_propd6}
	\parbox{2.6cm}{\vspace{0.cm}\includegraphics[width=2.6cm]{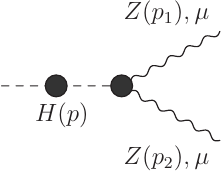}}
	= \frac{1}{p^2 -m_{H}^2} g^{SM}_{HZZ}\,,
\end{equation}
i.e. the momentum-dependent dimension-6 modifications cancel at the leading order in the $1/\Lambda^2$ expansion. This is because the modification correlates the $Q_{\square \Phi}$ modifications between the broken and unbroken phases~\cite{Englert:2019zmt}. In the linearised approximation, this also removes sensitivity in weak boson fusion signatures that are traditionally tell-tale signatures of electroweak modifications that temper with unitarity. This cancellation is therefore unique to the way how electroweak symmetry is broken. If the Higgs boson has a singlet component that feels the presence of the HEFT-like operator (see also~\cite{Anisha:2023ltp})
\begin{equation}
{\cal{O}}_{\square \square} = {a_{\square \square }\over v^2}  \square H \square H\,,
\end{equation}
the cancellation detailed above will {\emph{not}} occur. We can therefore expect non-trivial momentum dependencies in HEFT that are not predicted from SMEFT correlations, which can be exploited to set constraints (see also Ref.~\cite{Brivio:2014pfa}).

\begin{table}[b!]
	\centering
	\renewcommand{\arraystretch}{2.0}
	\small{	\begin{tabular}{|c|c|}
			\hline
			$\mathcal{O}_{HBB}$ 
			&$-a_{HBB}  \; g'^2\frac{H}{v} \text{Tr}\Big[ (B_{\mu \nu}\frac{\tau^3}{2}) (B^{\mu \nu}\frac{\tau^3}{2})\Big]$ \\
			\hline
			$\mathcal{O}_{HWW}$ 
			&$ -a_{HWW}\;g_{W}^2 \frac{H}{v} \text{Tr}\Big[ (W^{a}_{\mu \nu} \frac{\tau^{a}}{2}) (W^{a\mu \nu}\frac{\tau^{a}}{2})\Big]$ \\
			\hline
			$\mathcal{O}_{\square \boldsymbol{\mathcal{V}} \boldsymbol{\mathcal{V}} }$ 
			& $a_{\square \boldsymbol{\mathcal{V}} \boldsymbol{\mathcal{V}}}  \frac{\square H}{v} \text{Tr}\Big[ \boldsymbol{\mathcal{V}}_{\mu} \boldsymbol{\mathcal{V}}^{\mu}\Big]$ \\
			\hline
			$\mathcal{O}_{H0}$&
			$a_{H0} (M_Z^2 - M_W^2) \frac{H}{v} \text{Tr}\Big[\U \tau^3 \U^{\dagger} \boldsymbol{\mathcal{V}}_{\mu} \Big] \text{Tr}\Big[\U \tau^3 \U^{\dagger} \boldsymbol{\mathcal{V}}_{\mu} \Big]$  \\
			\hline
			$\mathcal{O}_{H1}$&
			$a_{H1} \; g' g_{W}\frac{H}{v} \text{Tr}\Big[\U B_{\mu\nu} \frac{\tau^3}{2} \U^{\dagger} W^{a}_{\mu\nu}\frac{\tau^{a}}{2}\Big]$  \\
			\hline
			$\mathcal{O}_{H8}$&
			$-\frac{a_{H8}}{4} \; g_{W}^2 \frac{H}{v} \text{Tr}\Big[\U \tau^3 \U^{\dagger}  W^{a}_{\mu\nu}\frac{\tau^{a}}{2}\Big] \text{Tr}\Big[\U \tau^3 \U^{\dagger}  W^{a}_{\mu\nu}\frac{\tau^{a}}{2}\Big]$  \\
			\hline
			$\mathcal{O}_{H11}$&
			$a_{H11} \frac{ H}{v} \text{Tr}\Big[ \mathcal{D}_{\mu} \boldsymbol{\mathcal{V}}^{\mu} \mathcal{D}_{\nu}\boldsymbol{\mathcal{V}}^{\nu}\Big]$ \\
			\hline
			$\mathcal{O}_{H13}$&
			$-\frac{a_{H13}}{2} \; \frac{H}{v} \text{Tr}\Big[\U \tau^3 \U^{\dagger}  \mathcal{D}_{\mu} \boldsymbol{\mathcal{V}}_{\nu}\Big] \text{Tr}\Big[\U \tau^3 \U^{\dagger}  \mathcal{D}^{\mu} \boldsymbol{\mathcal{V}}^{\nu}\Big]$  \\
			\hline
			$\mathcal{O}_{d1}$ 
			& $ i a_{d1} \; g' \frac{\partial^{\nu} H}{v} \text{Tr}\Big[ \U B_{\mu \nu} \frac{\tau^3}{2} \U^{\dagger} \boldsymbol{\mathcal{V}}^{\mu}\Big]$ \\
			\hline
			$\mathcal{O}_{d2}$ 
			& $ i a_{d2} \; g_{W} \frac{\partial^{\nu} H}{v} \text{Tr}\Big[ W^{a}_{\mu \nu}\frac{\tau^{a}}{2} \boldsymbol{\mathcal{V}}^{\mu}\Big]$ \\
			\hline
			$\mathcal{O}_{d3}$ 
			& $ a_{d3}  \frac{\partial^{\nu} H}{v} \text{Tr}\Big[  \boldsymbol{\mathcal{V}}^{\mu} \mathcal{D}_{\mu}\boldsymbol{\mathcal{V}}^{\mu}\Big]$ \\
			\hline
			$\mathcal{O}_{d4}$ 
			& $ a_{d4} \; g_{W} \frac{\partial^{\nu} H}{v}  \text{Tr}\Big[\U \tau^3 \U^{\dagger}  W^{a}_{\mu\nu}\frac{\tau^{a}}{2}\Big] \text{Tr}\Big[\U \tau^3 \U^{\dagger} \boldsymbol{\mathcal{V}}^{\mu} \Big]$ \\
			\hline
			$\mathcal{O}_{\square 0}$&
			$a_{\square0} \frac{(M_Z^2 - M_W^2)}{v^2} \frac{\square H}{v} \text{Tr}\Big[\U \tau^3 \U^{\dagger} \boldsymbol{\mathcal{V}}_{\mu} \Big] \text{Tr}\Big[\U \tau^3 \U^{\dagger} \boldsymbol{\mathcal{V}}_{\mu} \Big]$  \\
			\hline
			$\mathcal{O}_{\square \square}$ 
			& $a_{\square \square }  \frac{\square H \square H}{v^2}$ \\
			\hline
	\end{tabular}}
	\caption{Relevant HEFT operators $\mathcal{O}_{i}$  with $a_{i}$ being the corresponding HEFT coefficients. $\boldsymbol{\mathcal{V}}_{\mu}= (D_{\mu}\U)\U^{\dagger}$ and $\mathcal{D}_{\mu}\boldsymbol{\mathcal{V}}^{\mu}= \partial_{\mu}\boldsymbol{\mathcal{V}}^{\mu} + i [g_{W}W^{a}_{\mu}{\tau^{a}}/{2},\boldsymbol{\mathcal{V}}^{\mu}] $.} 
	\label{tab:operators}
\end{table}  

\section{HEFT interactions, SMEFT relations, amplitudes}
\label{sec:chiralEFT}
\subsection{HEFT interactions}
The leading order HEFT Lagrangian relevant for our study is given by
\begin{subequations}
	\label{eq:heftlo}
	\begin{multline}
		\mathcal{L} = - \frac{1}{4} W^{a}_{\mu \nu} W^{a\mu \nu} -  \frac{1}{4} B_{\mu \nu} B^{\mu \nu}\\ + {\cal{L}}_{\text{ferm}}+ \mathcal{L}_{\text{Yuk}} + \frac{ v^2}{4} \mathcal{F}_{H} \,\text{Tr}[D_{\mu} \U^{\dagger} D^{\mu} \U]    \\
		+ \frac{1}{2} \partial_{\mu} H \partial^{\mu} H - V(H) \,,
	\end{multline}
where, the matrix $\U  = \exp\left({i \pi^{a} \tau^{a}/v}\right)$ defines the Goldstone bosons $\pi^{a}$ in a  non-linear parametrisation with $\tau^{a}$ being the Pauli matrices for $a= 1,2,3$ and its covariant derivative is written as
\begin{equation}
	D_{\mu}\U = \partial_{\mu} \U + i g_{W} (W^{a}_{\mu} \tau^{a} /2) \; \U -i g' \U B_{\mu}\tau^{3}/2 \,.
\end{equation}
 $\mathcal{F}_{H}$ is the flare function giving the  Higgs interactions with gauge and Goldstone bosons and is given as
\begin{equation}
	\mathcal{F}_{H} = \Big(1+ 2(1+\zeta_{1})\frac{H}{v} + (1+\zeta_{2}) \Big(\frac{H}{v}\Big)^2 + ... \Big)\,.
\end{equation}
The couplings of Higgs boson with fermions are  
	\begin{equation}
	\mathcal{L}_{\text{Yuk}} = - \frac{v}{\sqrt{2}} \begin{pmatrix}
		\bar{u}^i_{L}  & \bar{d}^i_{L}
	\end{pmatrix} \U  \begin{pmatrix}
		\mathcal{Y}^{u}_{ij} u^{j}_{R}  \\ \mathcal{Y}^{d}_{ij} d^{j}_{R} 
	\end{pmatrix} + \text{h.c.}\,,
\end{equation}
and $\mathcal{Y}^{f}_{ij} $ is the function similar to $\mathcal{F}_{H}$ denoting the Higgs fermion interactions as
\begin{equation}
	\mathcal{Y}^{f}_{ij} = y^{f}_{ij} \Big(1+ (1+a_{1f})\,\frac{H}{v}+ ...\Big)\,.
\end{equation}
\end{subequations} 
The light quarks and leptons are neglected throughout our work. Here, $y^{f}_{ij}$ are the Yukawa couplings directly related to the mass terms. This leading-order HEFT Lagrangian can be extended by the chiral dimension four operators tabled in Tab.~\ref{tab:operators} reflect generic BSM connected to a custodial singlet nature of the Higgs boson. These interactions will be sourced at one-loop order from the leading order (chiral dimension 2) Lagrangian, Eq.~\eqref{eq:heftlo}, see \cite{Herrero:2021iqt,Herrero:2020dtv,Anisha:2022ctm,Herrero:2022krh} and can have significant implications for phenomenological observations~\cite{Davila:2023fkk,Englert:2023uug}. For concrete matching computations related to HEFT, see the recent Refs.~\cite{Buchalla:2016bse, Arco:2023sac,Dawson:2023oce}.

In our work, we further assume that the electroweak precision constraints are not violated and the oblique $S$, $T$ and $U$ parameters~\cite{Peskin:1991sw} are related to the following chiral dimension four operators~\cite{Brivio:2016fzo}
 \begin{equation}
 	S \propto a_{H1} , \hspace{0.5cm} T \propto a_{H0}, \hspace{0.5cm} U \propto a_{H8}\,.
 \end{equation}  
Thus, these above-mentioned operators are predominantly constrained from electroweak precision data, and to explore the sensitivity of the off-shell measurement to Higgs non-linearity, we set these coefficients to zero. (The potential shortfalls of such assumptions in the context of global fits and SMEFT have been highlighted in Ref.~\cite{Berthier:2015oma,Bellafronte:2023amz}.)

\subsection{SMEFT from HEFT} \label{subsec:SMEFTrules}
\begin{figure}[!t]
	\includegraphics[width=0.485\textwidth]{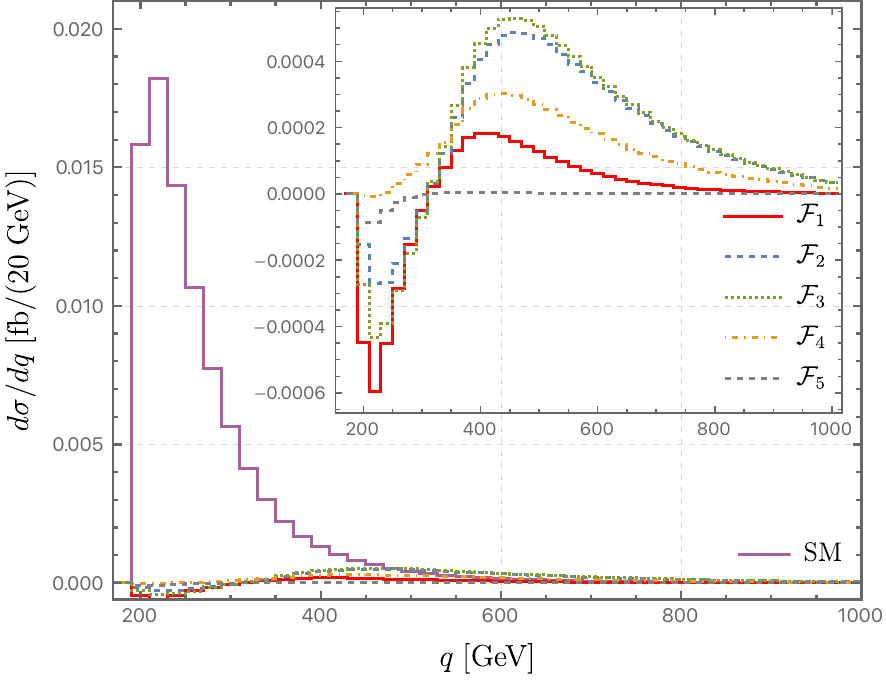}
	\caption{ \label{fig:histogram_zz} Off-shell Higgs momentum distributions for $H\to ZZ$ for SM and with the different form factors ${\cal{F}}_{i}=1$ shown in the inset.}
\end{figure}

Some of the operators listed in Table~\ref{tab:operators} are related to the following dimension six SMEFT operators in the Warsaw basis~\cite{Grzadkowski:2010es}
\begin{equation}
	\begin{split}
		Q_{\Phi B} & = \frac{C_{\Phi B}}{\Lambda^2} \Phi^{\dagger}\Phi \; B_{\mu \nu} B^{\mu \nu}\,, \\
		Q_{\Phi W} & =  \frac{C_{\Phi W}}{\Lambda^2} \Phi^{\dagger}\Phi \; W^{a}_{\mu \nu} W^{a,\mu \nu}\,, \\
		Q_{t\Phi}  &= \frac{C_{t\Phi}}{\Lambda^2} \big(\Phi^{\dagger}\Phi \;  ( \bar{Q}\,t\,\tilde{\Phi})+ \text{h.c.}\big )\,. 
	\end{split}
\end{equation}\\
Here $Q$'s are the SMEFT operators and $C$'s are the corresponding Wilson Coefficients (WCs) with $\Lambda$ being the cut-off scale. The translation rules between the non-linear and linear coefficients are 
\begin{equation} \label{eq:SMEFTrules}
\begin{split}
	a_{HBB} &=  -2  \frac{v^2}{ g'^2} \, \frac{ C_{\Phi B}}{\Lambda^2}\,, \\
	a_{HWW} &=  -2 \frac{v^2}{ g_{W}^2}\, \frac{C_{\Phi W}}{\Lambda^2}\,, \\
	a_{1t} &=- \frac{v^3}{\sqrt{2}M_{t}} \frac{C_{t\Phi}}{\Lambda^2}\,.
\end{split}
\end{equation}
We need to go to the higher mass dimension (larger than $d=6$) to obtain the correspondence of the other HEFT operators.
\begin{widetext} 
\subsection{Amplitudes and Implementation}
We implement the EFT corrections using form factors~\cite{Hagiwara:1986vm}. Concretely, we extract the independent Lorentz structures contributing to the Higgs amplitudes decay 
amplitudes \hbox{$H\to VV$} after performing on-shell renormalisation as described above. The relevant Higgs off-shell $H\to VV$ corrections, including the expanded Higg boson propagator, are then given by
\newcommand{\ep}{\varepsilon}
\begin{subequations}
\begin{multline}
\label{eq:form_zz}
 i\widetilde{\Gamma}_{HVV} 
=   -{e^2 m_t\over 2\cw^2\sw^2}
{1\over q^2-M_H^2 + i\Gamma_H M_H}  \\
 \bigg\{\left[ 
( 1 + {\cal{F}}_1 ) 
+ {{\cal{F}}_2 \over v^2}  (q^2-p_1^2-p_2^2) 
+ {{\cal{F}}_3 \over v^2} q^2   
+  {{\cal{F}}_5 \over v^2} {M_H^2 \over q^2 - M_H^2 + i\Gamma_H M_H}
\right] [\ep^\ast(p_1) \cdot \ep^{\ast}(p_2)]
+ {{\cal{F}}_4 \over v^2}  [\ep^\ast(p_1) \cdot p_2][\ep^\ast(p_2) \cdot p_1]\bigg\}\,,
\end{multline}
where we have included the Yukawa coupling alongside its corrections arising from $t\bar t \to H(q)$, with $q= p_1+p_2$ ($M_H,\Gamma_H$ denote the Higgs boson mass and width, respectively). Matching the Lorentz structures to the HEFT coefficients (which contain a SMEFT limit), we find
\begin{align}
{\cal{F}}_1 &= a_{1t} + 2 {a_{\Box\Box}} {M_H^2\over v^2} + \zeta_1 \,,\\
{\cal{F}}_2 &= a_{H13} + 2 a_{HBB} \sw^4 + 2 a_{HWW} \cw^4\,, \\
{\cal{F}}_3 &= a_{\Box BB}  - 2 {a_{\Box\Box}} + a_{d2} + 2 a_{d4} + {e^2 \over \cw^2} a_{\Box 0} - (a_{d1}-a_{d2}-2a_{d4}) \sw^2\,, \\
{{\cal{F}}_5}  &= 2 a_{\Box\Box}\,,\\
{{\cal{F}}_4\over 2}  &= (a_{d1} +4 a_{HWW} )\sw^2 - 2 a_{HWW} - (a_{d2} +2 a_{d4} ) \cw^2 - 2 (a_{HBB} + a_{HWW}) \sw^4 \,, 
\end{align}
\end{subequations}
\end{widetext}
for $H\to ZZ$. A similar decomposition holds for $H\to WW$. ${\cal{F}}_5$ arises from corrections to the Higgs propagator, Eq.~\eqref{eq:higgsprop}. As this is obtained from the 2-point vertex function (i.e. the inverse propagator), the corrections related to ${\cal{F}}_5$ are intrinsically dependent on the truncation at chiral dimension 4 (equivalent to dimension 6 for the SMEFT identification). To some extent ${\cal{F}}_5$ therefore probes a truncation scheme dependence, in particular in the off-shell regime where the LHC experiments perform their measurement $q \gtrsim 350~\text{GeV}$. Sensitivity to ${\cal{F}}_5$ is comparably suppressed to the other ${\cal{F}}_i$, and we can therefore trust the truncation as detailed here, Fig.~\ref{fig:histogram_zz}.
The amplitudes and their corrections relative to the SM have been implemented using {\sc{Vbfnlo}}~\cite{Arnold:2008rz,Baglio:2014uba} (including cross-checks against the results of \cite{Campbell:2013wga,Campbell:2013una}). For this study we limit ourselves to linear new physics contribution, i.e. the differential cross sections are truncated at linear order of the HEFT coefficients.

The phenomenological significance of the off-shell measurement lies in its correlation with on-shell Higgs quantities. For instance, Eq.~\eqref{eq:form_zz} introduces modifications to the Higgs branching to vector bosons. These changes are~\cite{Rizzo:1980gz,Keung:1984hn,Dawson:2018dcd,Dawson:2018pyl}, for $H(p_H) \to f(p_1) f(p_2) V(p_3)$
\begin{subequations}
\begin{multline}
\label{eq:offshellhvv}
\Gamma(H\to VV^\ast) =\\ \int_{0}^{(M_H - M_V)^2} \hbox{d} m_{12}^2 \int_{m_{23,\text{l}}^2}^{m_{23,\text{u}}^2} \hbox{d} m_{23}^2\, \dfrac{{|\overline{{\cal{M}}}|}^2}{ (2\pi)^3\, 32 \,M_H^3}\,,
\end{multline}
with $m_{ij}^2 = (p_i + p_j)^2$ and, assuming massless fermions $f$
\begin{equation}
m_{12}^2 + m_{23}^2 + m_{13}^2 = M_H^2 + M_V^2\,.
\end{equation}
The limits of the $m_{23}^2$ integration are 
\begin{multline}
\lambda(m_{12},M_H,M_V) =\\ m_{12}^4 - 2 m_{12}^2 (M_H^2 + M_V^2) + (M_H^2 - M_V^2)^2\,, 
\end{multline}
with
\begin{equation}
2m_{23,\text{l},\text{u}}^2=M_H^2 + M_V^2 - m_{12}^2 \mp \sqrt{\lambda(m_{12},M_H,M_V)}\,.
\end{equation}
\end{subequations}
For the results in the next section, again, we limit ourselves to the linear order in the HEFT coefficients, i.e. the spin-summed/averaged matrix element $|\overline{{\cal{M}}}|^2$ only contains interactions $\sim a_i$ given in Tab.~\ref{tab:operators}. Turning to the $H\to \gamma\gamma $ decay width, the scattering amplitude is therefore given as 
\begin{equation}
	\mathcal{M}= |\mathcal{M}_{\text{1-loop}}|^2 + 2\,\text{Re}\big(\mathcal{M}_{\text{HEFT}}^{*}\mathcal{M}_{\text{1-loop}}\big)\,,
\end{equation}
where $\mathcal{M}_{\text{1-loop}}$ is the one-loop amplitude calculated with LO Lagrangian given in Eq.~\eqref{eq:heftlo} (includes parameters $\zeta_{1}$ and $a_{1t}$) and $\mathcal{M}_{\text{HEFT}}$ is the amplitude generated with HEFT operators given in Table~\ref{tab:operators}. This gives rise to the $H\to \gamma \gamma$ modification and, similarly, $H\to \gamma Z, gg$ can be derived. We do not repeat this here but refer the interested readers to the existing literature~\cite{Herrero:2020dtv,Herrero:2022krh,Anisha:2022ctm}, which we have cross-checked our results against. The detailed expressions of the Higgs decay widths and the contribution to the total Higgs width are listed in appendix~\ref{appendix:SS}.

\begin{figure*}[!t]
	\centering
	\subfloat[]
	{\includegraphics[width=0.35\textwidth]{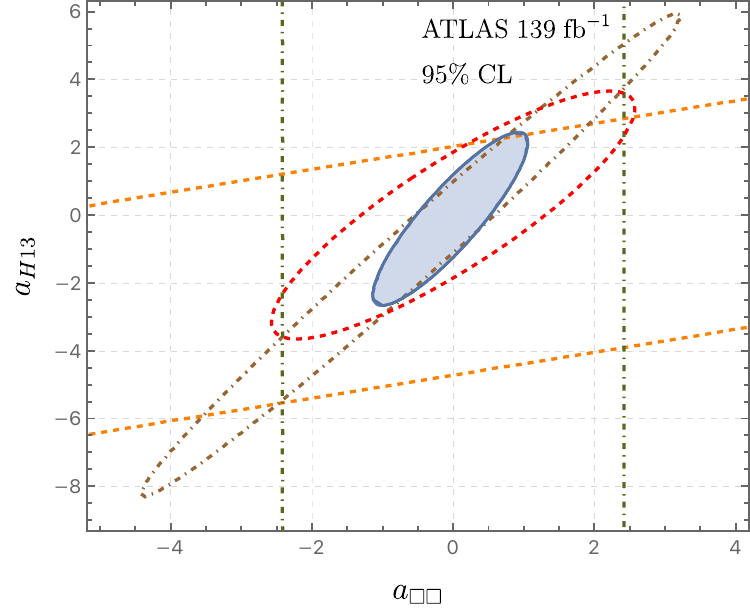}}~
	\subfloat[]
	{\includegraphics[width=0.35\textwidth]{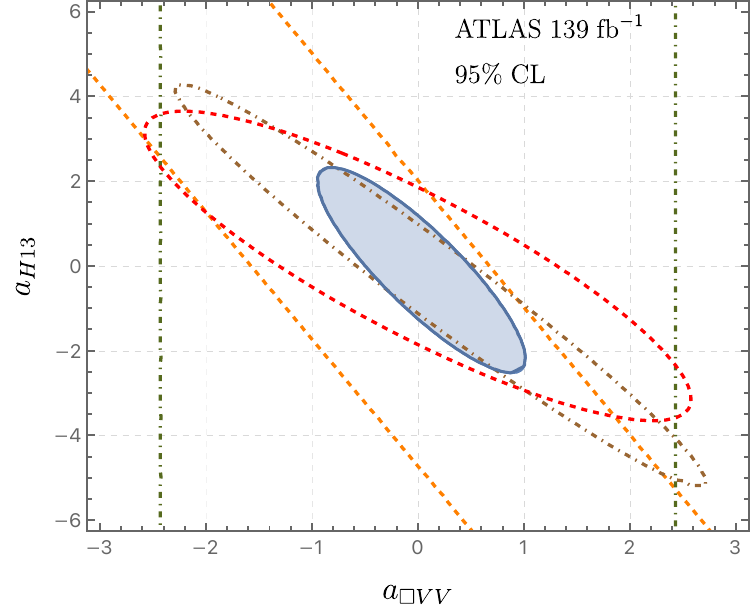}}~
	{\includegraphics[width=0.2\textwidth]{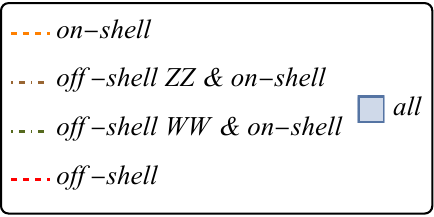}}
	\caption{ \label{fig:LHCbounds} 95\% CL contours obtained for the HEFT coefficients using the ATLAS 139 $\text {fb}^{-1}$ data. These regions are obtained after profiling over SMEFT WCs. }
\end{figure*}

\begin{figure*}[!t]
	\centering
	\subfloat[]
	{\includegraphics[width=0.35\textwidth]{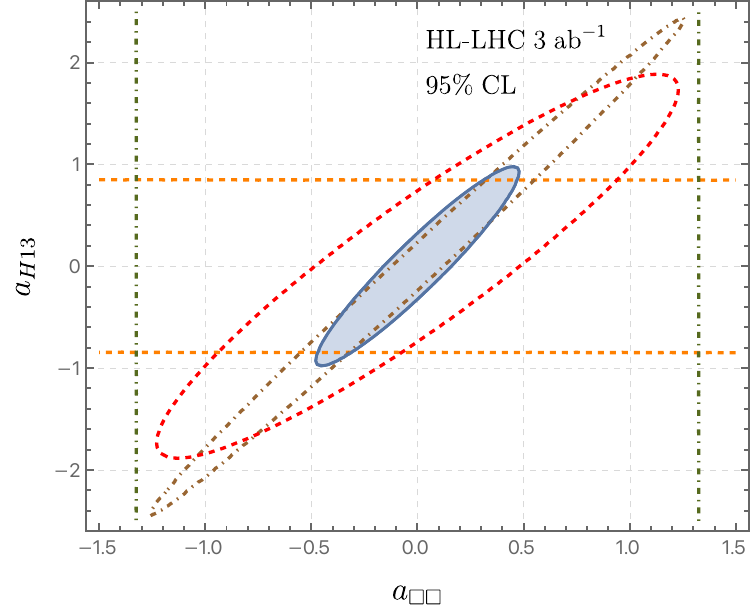}}~
	\subfloat[]
	{\includegraphics[width=0.35\textwidth]{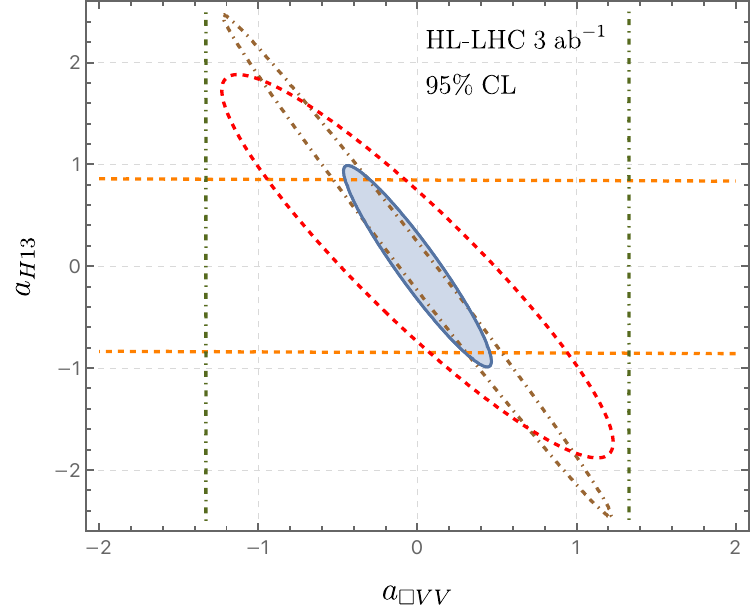}}~
    {\includegraphics[width=0.2\textwidth]{legend.pdf}}
	\caption{ \label{fig:HLLHC} Two-dimensional allowed 95\% CL regions for the HEFT coefficients  using the $\chi^2$ analysis from the HL-LHC projected data. The HL-LHC extrapolations do not include correlations.  $\chi^2_{\text{SM}}=0$ results in the tilt of the on-shell-only constraints compared to Fig.~\ref{fig:LHCbounds}.}
\end{figure*}

\section{Off-shellness as a probe of Non-Linearity}
\label{sec:offshell}
With amplitudes and HEFT-SMEFT relations in place, we can now turn to a quantitative estimate of the sensitivity of the off-shell measurement to Higgs boson non-linearity. To this end, we assume 85\% efficiency of the $4\ell$ sample after selection including a flat 29\% systematic following~\cite{Englert:2015hrx,CMS-PAS-HIG-21-019}. To gain a qualitative statistical understanding of expected constraints, we include these as binned $\chi^2$ test statistic
\begin{equation}
	\chi^2 = \sum_i {(N_i - N_i^\text{SM})^2\over \sigma^2_{i,\text{syst}} + \sigma^2_{i,\text{stat}}}\,,
\end{equation}
where $N_i$ denotes the entries in the $i$th bin of the off-shell Higgs momentum distribution with HEFT contributions. For $ZZ$ processes, we have applied a cut of $q>300$~GeV on the invariant $ZZ$ mass. The variable $N_i^{\mathrm{SM}}$ denotes the SM expectation in bin $i$, and $\sigma_{i,\mathrm{stat}}$ and $\sigma_{i,\mathrm{syst}}$ are the statistical and systematic uncertainties, respectively. We approximate these with $\sigma_{i,\mathrm{stat}} = \sqrt{N_i^{\mathrm{SM}}}$ and $\sigma_{i,\mathrm{syst}} = 0.29 N_i^{\mathrm{SM}}$. The latter term includes uncertainties in background processes~\cite{CMS-PAS-HIG-21-019}, for example from continuum production of the form $q\bar{q} \to 4 \ell$.

To include the constraints from on-shell Higgs data, using the modifications in the Higgs partial decay widths, the total Higgs decay width is calculated (expressions are listed in appendix~\ref{appendix:SS}). With these, the modified Higgs branching ratios are given as
\begin{equation}
	 \frac{\text{BR}^{\text{HEFT}}(H\to X)} { {\text{BR}}^{\text{SM}}(H\to X) }
  =   \frac{\Gamma^{\text{HEFT}}(H\to X)} {\Gamma^{\text{SM}}(H\to X) } 
  \frac{ \Gamma^{\text{SM}}_{H}}  { \Gamma^{\text{HEFT}}_{H}}	
\end{equation}
Here $X$ denotes $\gamma \gamma$, $VV^{*}$, and $\gamma Z$.\footnote{The modifications to the $b\bar{b}$ partial width are ignored in our analysis. For the total decay width, in addition to the corrections to these partial decay widths, we also scale the rest of the decay widths with the Higgs field redefinition factor. We rescale the leading order results to reproduce the SM expectation of~\cite{LHCHiggsCrossSectionWorkingGroup:2011wcg}.} Combining these branching ratios with the change in the production cross-section of gluon-gluon fusion and assuming narrow width approximation, the modified signal strengths are obtained as
\begin{equation}
	\mu^{X}_{\text{ggF}}= \frac{[\sigma_{\text{ggF}} \; \text{BR}(H\to X)]^{\text{HEFT}}}{[\sigma_{\text{ggF}} \; \text{BR}(H\to X)]^{\text{SM}}}
\end{equation}
The on-shell $\chi^2$ statistic constructed using the signal strengths from 139 fb$^{-1}$ data~\cite{ATLAS-CONF-2021-053} is given as
\begin{multline}\label{eq:chi2}
	\chi^2_{\text{on-shell}} \\= \sum_{i,j = 1 }^{\text{data}}\big(\mu_{i, \text{exp}}- \mu_{i,\text{th}}\big) (V_{ij})^{-1} \big(\mu_{j, \text{exp}}- \mu_{j,\text{th}}).
\end{multline}
Here, $\mu_{\text{th}}$ are the theory expressions incorporating the effects of HEFT operators. $\mu_{\text{exp}}$ are the central values of the experimental measurements and the covariance matrix $V = \rho_{ij} \sigma_{i} \sigma_{j}$ with $\rho$ being the correlation matrix and $\sigma$ the uncertainties. To obtain an estimate of the improvements expected in future HL-LHC runs, we take the projections to 3 ab$^{-1}$ from the recent HL-LHC analysis of Ref.~\cite{Dainese:2703572}. The details of the signal strengths measurements are provided in appendix~\ref{appendix:SS} for completeness. 

\begin{table}[b]
	\centering
	\renewcommand{\arraystretch}{1.2}
	\begin{tabular}{|c|c|c|}
		\hline 
		\multirow{2}{*}{{ Datasets}}	&   \multicolumn{2}{c|}{$\Delta{\chi^2}$ values} \\ 
		\cline{2-3} & ATLAS 139 fb$^{-1}$ & HL-LHC 3 ab$^{-1}$  \\
		\hline 
		{\it on-shell} &  3.84 &  	5.99  \\
		{\it off-shell ZZ} \&~{\it on-shell} & 49.80 & 50.99 \\
		{\it off-shell WW} \&~{\it on-shell} & 54.57 & 55.76\\
		{\it off-shell} & 90.53 & 90.53\\
		{\it all} & 93.95 & 95.08\\
		\hline
	\end{tabular}
	\caption{$\Delta\chi^2$ constraints used to obtain 95\% CL regions depending on the total degrees of freedom. }
	\label{tab:chi2_constraints}
\end{table}

We can now turn to our results. To make a qualifying statement about linearity vs. non-linearity, we profile the SMEFT contributions detailed in Sec.~\ref{subsec:SMEFTrules}.\footnote{Here `profiling' refers to minimising the $\chi^2$ simultaneously using HEFT operators ($a_{1t}$, $a_{HWW}$, $a_{HBB}$) which are related to SMEFT operators, see Eq.~\eqref{eq:SMEFTrules}.} This way, we obtain a statistical measure of non-linearity as expressed through the sensitivity to the corresponding HEFT coefficients. While interactions like $a_{\Box \Box}$ enter as a uniform coupling rescaling, it is predominantly probed in the off-shell region. The on-shell region therefore probes predominantly on-shell related quantities whereas the energy-related scaling of HEFT vs. SMEFT is visibly expressed by, e.g., $a_{\Box \Box}$. We have also included the $H\to WW $ off-shell region; due to its less straightforward final state phenomenology, this mode has received less attention compared to the fully reconstructible $H\to ZZ \to 4\ell$ final states. (Some of the selection criteria, see e.g. \cite{Campbell:2013wga}, also lead to a significant reduction of the off-shell region). Assuming the same (perhaps optimistic) systematic uncertainties and efficiencies as for $H\to 4\ell$, we do not find a significant information gain when including $H\to WW$ through the transverse mass observable. 

We combine the on and off-shell contributions to a total $\chi^2$ statistic to understand the on-shell vs off-shell effects, treating these phase space regions as statistically uncorrelated.  The total number of data points upon combining on-shell signal strength data with the off-shell binned data is 76 (75) for HL-LHC (LHC). To obtain the 95\% CL regions, we constrain the $\chi^2 $ statistic with the $\Delta\chi^2$ values obtained from the degrees of freedom, i.e. (number of data points - number of parameters). After profiling over the SMEFT directions, for our {\it all} dataset (i.e. then on-shell/off-shell combination), the strongest bounds can be imposed on the operators parameterised by $a_{\Box \Box}$, $a_{H13}$, $a_{\square VV}$ and $a_{d4}$ (when these are considered in isolation). Using pairwise combinations of these operators, we show the two-dimensional parameter space allowed at 95\% CL with both 139 fb$^{-1}$ and 3 ab$^{-1}$ data in Figs.~\ref{fig:LHCbounds} and \ref{fig:HLLHC}. The constraints used for the regions are outlined in Table~\ref{tab:chi2_constraints} for different combinations of datasets considered in the analysis (again the remaining HEFT directions are assumed to be zero)\footnote{The allowed parameter space is extremely large after profiling over all HEFT operators}.  The most stringent impact indeed arises from the inclusion of the off-shell measurements, shown by the red dashed contour (predominantly $H \to ZZ$) in Fig.~\ref{fig:LHCbounds}.

The HL-LHC extrapolation rests on the YR18 systematic uncertainties which include a scaling of systematic uncertainties with the root of the collected luminosity. This is relatively pessimistic and it is therefore not unlikely that systematics become under much better control than what can be forecast now. In such a situation we can expect stronger limits on the BSM coupling space across many relevant Higgs production and decay modes, beyond the off-shell measurement detailed here. Of course, the relatively small data set that we have considered in this proof-of-principle analysis is not large enough to control all relevant HEFT Higgs interactions and a global fit of the discussed modes will have little sensitivity. However, the inclusion of weak boson fusion and multi-Higgs final states will add further sensitivity. Weak boson fusion appears to be particularly motivated as our discussion will directly generalise to $WW$ scattering. We leave this, as well as a more global fit, for future work.

\begin{table*}[!t]
	\centering
	\renewcommand{\arraystretch}{1.8}
	\begin{tabular}{|c|c|ccc|c|}
		\hline 
		\multirow{2}{*}{{ Observables}} &  \multicolumn{4}{c|}{ATLAS Run 2 data~\cite{ATLAS-CONF-2021-053}	} & \multirow{2}{*}{{HL-LHC uncertainties~\cite{Dainese:2703572} }} \\ 
		\cline{2-5} & Measurements & \multicolumn{3}{c|}{Correlations} & \\
		\hline 
		$\mu^{\gamma\gamma}_{ggF}$ 	& $1.02^{+0.11}_{-0.11}$  & $1 \;$   & $0.05$& $0.09$ & $\pm 0.36$ \\
		$\mu^{ZZ}_{ggF}$  			& $0.95^{+0.11}_{-0.11}$ &  & $1$& $0.1$  & $\pm 0.039$  \\
		$\mu^{WW}_{ggF}$ 	& $1.13^{+0.13}_{-0.12}$ &  &  &$1$ & $\pm 0.043$  \\ 
		\cline{1-5}
		$\mu^{Z\gamma}_{ggF}$ & \multicolumn{4}{c|}{}& $\pm 0.33$ \\
		\hline
	\end{tabular}
	\caption{Details of the signal strength measurements used in the $\chi^2_{\text{on-shell}}$. Columns~2 and 3 list the ATLAS 139~fb$^{-1}$ data. Column~4 lists the projections used for HL-LHC 3 ab$^{-1}$.  }
	\label{tab:exp_SS}
\end{table*}

\section{Conclusions}
\label{sec:conc}
The non-decoupling Higgs contribution in $gg\to VV$ production is a versatile tool to gain sensitivity to new physics beyond the Standard Model. Any deviation from expected SM coupling patterns filters through to modified tail contributions as a consequence of the interplay of absorptive amplitude parts that, in the SM, are determined by unitarity and, hence, renormalisablilty~\cite{Kauer:2012hd} from various angles. The most prevailing of these interpretations is the phrasing of off-shell constraints as on-shell measurements, which has brought this measurement to the fame it deserves~\cite{Caola:2013yja}. Such directions of interpretation rest on limiting assumptions~\cite{Englert:2014ffa} which suggest alternative ways of reporting outcomes of the measurement. To entice ATLAS and CMS to consider different avenues of interpretation, in this work, we have analysed the off-shell vs. on-shell correlation as a probe of Higgs boson non-linearity. Analyses that aim to distinguish linear from non-linear Higgs EFT modifications are typically focussed on a comparison of Higgs multiplicities~\cite{Gomez-Ambrosio:2022why,Bhardwaj:2023ufl,Delgado:2023ynh} (see also the recent \cite{Stylianou:2023xit,Papaefstathiou:2023uum}). In this exploratory study, we have shown that $gg\to VV$ straddles dual roles of fingerprinting unitarity departures {\emph{as well as}} deviations from SMEFT attributed to the propagation of the Higgs boson. These implications generalise to weak boson fusion where we can expect similar patterns in a HEFT vs. SMEFT comparison. 

Of course, we can always consider additional operators, whether they appear as part of a higher-dimensional SMEFT contribution, couplings of higher chiral dimension, or as part of a plethora of operators in a global fit. In this work we have limited ourselves to top-related interactions. Contact interactions $\sim ggH$ have not been considered, but it is known that these can be separated from top-mediated processes by resolving the top loop via $H+\text{jet}$ production~\cite{Banfi:2013yoa,Grojean:2013nya}. By profiling the SMEFT directions, we have obtained a statistical estimate of the constraints on non-SMEFT interactions that can be obtained, and this shows promise for the inclusion in a more comprehensive analysis, which we leave for future work.

\bigskip

\noindent {\bf{Acknowledgements}} --- 
A is funded by the Leverhulme Trust under RPG-2021-031. CE is supported by the UK Science and Technology Facilities Council (STFC) under grant ST/X000605/1 and the Leverhulme Trust under RPG-2021-031. RK is supported by the Helmholtz Association under the contract W2/W3-123. MS is supported by the STFC under grant ST/P001246/1. \\

\appendix 
\section{Higgs boson decay widths and LHC signal strength constraints}\label{appendix:SS}
The contribution of HEFT operators to the Higgs decay widths relative to the SM are listed below. The contribution to the gluon-gluon fusion production cross-section relative to SM is similar to $H \to gg$.
\begin{widetext}
\begin{align}
	\frac{\Gamma^{\text{HEFT}}(H \to ZZ)}{\Gamma^{\text{SM}}(H \to ZZ)} &= 1 + 0.059 \; a_{\square 0} - 0.99 \; a_{\square \square } + 
	0.99 \; a_{\square VV} - 0.09 \; a_{d1} \nonumber \\   &  + 0.32 \; a_{d2} + 0.64 \; a_{d4} + 
	0.17 \; a_{H13} + 0.007 \; a_{HBB} + 0.09 \; a_{HWW} + 2 \; \zeta_1\,, \nonumber \\
	\frac{\Gamma^{\text{HEFT}}(H \to WW)}{\Gamma^{\text{SM}}(H \to WW)} &=  1 - 0.99 \; a_{\square \square } + 0.99 \; a_{\square VV} + 
	0.40 \; a_{d2} + 0.05 \; a_{HWW} + 2 \; \zeta_1\,,\\
	\frac{\Gamma^{\text{HEFT}}(H \to \gamma \gamma )}{\Gamma^{\text{SM}}(H \to \gamma \gamma)} &=  1- 0.57 \; a_{1t} - 0.99 \; a_{\square \square} + 48.67 \; a_{HBB} + 
	48.67 \; a_{HWW} + 2.57 \; \zeta_1\,,\nonumber\\
	\frac{\Gamma^{\text{HEFT}}(H \to  \gamma Z)}{\Gamma^{\text{SM}}(H \to  \gamma Z)} &= 1- 0.12 \; a_{1t} - 0.99 \; a_{\square \square} + 16.26 \; a_{d1} + 
	16.26 \; a_{d2} \nonumber \\  &+ 32.52 \; a_{d4} - 14.43 \; a_{HBB} + 50.61 \; a_{HWW} + 2.12 \; \zeta_1\,,\nonumber \\
	\frac{\Gamma^{\text{HEFT}}(H \to gg )}{\Gamma^{\text{SM}}(H \to gg)}& =  1+ 2 \; a_{1t} - 0.99 \; a_{\square \square}\,. \nonumber
\end{align}
Using the above expressions, the HEFT contribution to the total Higgs decay width is 
\begin{align}
	\frac{\Gamma^{\text{HEFT}}_H}{\Gamma^{\text{SM}}_H } &= 1 + 0.17 \; a_{1t} + 0.001 \; \; a_{\square 0} - 
	0.99 \; \; a_{\square \square} + 0.24 \; a_{\square VV} \nonumber \\  & + 0.02 \; \; a_{d1} + 0.12 \; \; a_{d2} + 
	0.07 \; \; a_{d4} + 0.004 \; a_{H13} + 0.09 \; a_{HBB} + 0.20 \; a_{HWW} + 
	0.5 \; \zeta_{1} \,.
	\end{align}
\end{widetext}

The signal strength measurements used in the $\chi^2_{\text{onshell}}$ for 139 fb$^{-1}$ data along with correlation matrix are shown in Table~\ref{tab:exp_SS}. The HL-LHC 3 ab$^{-1}$ projections of the signal strengths are shown in column~4.
\bibliography{paper}

\end{document}